\documentclass[nonacm, sigconf]{acmart}
\AtBeginDocument{%
  }

\usepackage{cleveref}
\usepackage{tabularx, tabulary}
\usepackage[table]{xcolor}
\usepackage{xspace}
\usepackage{graphicx}
\usepackage{subcaption}
\usepackage{multirow}
\usepackage{tikz}

\newcommand{\sysname}{\textsc{S-HPLB}\xspace}
\newcommand{\phm}[1]{\vspace{.4em} \noindent\textbf{#1}\hspace{.5em}}

\newcommand*\circleb[1]{\tikz[baseline=(char.base)]{
    \node[shape=circle,fill,inner sep=.5pt](char){\textcolor{white}{\small #1}};}}

\begin{document}

\title{\sysname{}: Efficient LLM Attention Serving via Sparsity-Aware Head Parallelism Load Balance}

\author{Di Liu, Yifei Liu, Chen Chen$^{*}$, Zhibin Yu$^{\dagger}$, Xiaoyi Fan$^{\ddagger}$, Quan Chen, Minyi Guo}
\affiliation{%
  \institution{Shanghai Jiao Tong University, $^{\dagger}$Chinese Academy of Science, $^{\ddagger}$Shenzhen MSU-BIT University}
  \country{}
}
\thanks{$^{*}$Chen Chen is the corresponding author.}
\begin{abstract}
With the increasing volumes of Large Language Models (LLMs) and the expanding context lengths, attention computation has become a key performance bottleneck in LLM serving.
For fast attention computation, recent practices often parallelize the attention heads on multiple GPUs, and also widely adopt attention sparsification to reduce the computation amount---which selectively computes a subset of attention pairs under a preset sparsity budget. 
In this paper, we notice that attention heads of an LLM model often exhibit \emph{heterogeneous-yet-stable} sparsity elasticities, which motivates us to enforce head-adaptive sparsity budgets to attain better efficiency while preserving high inference quality.
Yet, from the system aspect, with heterogeneous sparsity levels, attention computation time on different heads would be inconsistent, yielding cross-GPU resource bubbles under head-parallel deployment.
To further minimize such bubbles, we propose a novel attention deployment strategy called \emph{Sparsity-aware Head-Parallel Load Balance} (S-HPLB).
Experiments on long-context benchmark show that, S-HPLB can achieve a $2.88\times$ improvement in average attention computation latency without quality degradation.

\end{abstract}

\maketitle

\section{Introduction}
\label{sec:intro}

Recent advances in large language models (LLMs) have enabled remarkable capabilities in natural language understanding and generation. 
Meanwhile, applications such as document understanding~\cite{Wang2024Leave}, long-form question answering~\cite{Li2025Long}, and code synthesis~\cite{jimenez2023swe,bairi2024codeplan} feed models with increasingly longer context inputs.
As both model size and context length grow, the computational demand of LLM inference---especially the \emph{attention} operation---escalates dramatically. 
For instance, when processing an input with a 128K context length using the Llama-3-8B model, attention operation accounts for over 70\% of the total time-to-first-token (TTFT)~\cite{liu2025retrievalattention}. 
Therefore, reducing the cost of attention operation is crucial for efficient LLM serving.

For efficient LLM attention serving, from the \emph{system} perspective, recent practices increasingly adopt distributed attention deployment~\cite{zhu2025megascale, wang2025step,Zheng2024Efficient, Kwon2023efficient}. 
On the one hand, to avoid cross-module resource interference, the attention module in each Transformer layer can be decoupled from the other modules (e.g., FFN), which is called attention-FFN disaggregation (AFD)~\cite{zhu2025megascale, wang2025step, xiao2025xdeepserve};
this renders it possible to optimize the attention efficiency independently.
On the other hand, to particularly speed up the attention module with expanded resources, leading LLM serving systems~\cite{Zheng2024Efficient, Kwon2023efficient} often evenly distribute the attention heads of an attention module on different GPUs, a deployment mechanism we call \emph{Head-parallelism} (or HP, which is a special form of the classical Tensor Parallelism). 

Meanwhile, from the \emph{algorithm} perspective, \emph{sparse attention} is a popular optimization technique in the literature~\cite{Zhang2024h2o, xiao2023efficient, tang2024quest, jiang2024minference}.
Specifically, the attention operation has inherent sparsity, meaning that each query vector essentially interacts with only a subset of the key vectors; sparse attention can thus alleviate the attention cost by bypassing the attention computation of insignificant query-key pairs. 
Moreover, in determining the number of tokens to select in attention computation, existing sparse-attention methods typically rely on a fixed token budget (i.e., adopting the top-$k$ policy where $k$ is the selected token number)~\cite{xiao2023efficient,jiang2024minference}, which is uniformly applied to all the attention heads.


However, in attaining high attention efficiency while still preserving the accuracy quality, we find that enforcing a uniform sparsity budget across different heads is far from optimal.  
In fact, our testbed measurements (\Cref{fig:moti_fixed_budget}) show that the attention heads of a Transformer layer usually exhibit diverse sparsity levels (i.e., with diverse relationship shapes between the overall attention weight and the sparsity budget).
In that case, enforcing a uniform token budget would lead to redundant computations on some (high-sparsity) heads, yet at the same time incur large attention loss on other (low-sparsity) heads.
Recently, some works~\cite{xu2025xattention, lin2025twilight} have proposed the top-$p$ method to implicitly handle such sparsity heterogeneity, which adaptively determines the token budget for each head---such that the total attention weight of its computed query-key pairs reaches a preset threshold $p$.
Nonetheless, such a top-$p$ method requires costly analysis of the attention map, and is often inaccurate in determining the attention participants~\cite{lin2025twilight}.
Besides, when attention heads have different sparsity budgets, the attention computation time on different GPUs would be inconsistent, suffering resource wastage due to the synchronization barriers. 
Motivated by the above discussions, in this paper we present \emph{Sparsity-aware Head Parallelism Load Balance} (\sysname{}), a system-algorithm co-designed mechanism to improve the end-to-end attention serving efficiency while maintaining high accuracy. 
From the algorithm aspect, we first observe that, 
despite the cross-head sparsity heterogeneity, on each head the sparsity pattern is however stable across different use cases that have diverse input lengths and tasks; this makes it possible to conduct per-head sparsity modeling in an offline manner.
Besides, given an expected sparsification ratio (i.e., the $k$ budget) and the profiled per-head sparsity pattern, we propose 
to better off the low-sparsity heads with cross-head budget shifting,
so as to enhance the accuracy performance without inflating the overall computation amount. 
Moreover, from the system aspect, 
with the algorithm-side sparsity budgets, we model the head deployment problem as a classical  multiway partitioning problem, and propose a greedy heuristic to minimize the resource wastage caused by cross-GPU load inconsistency. 

We evaluate both the accuracy and efficiency of \sysname{} on three leading open source LLMs, running the widely adopted RULER benchmark~\cite{hsieh2024ruler} on a server equipped with eight A100 GPUs.
Experimental results show that, compared to state-of-the-art sparse-attention algorithms, \sysname{} improves the model accuracy and attention latency by up to 5.34\% and 2.88$\times$, respectively. 
Further deep-dive studies demonstrate that, \sysname{} operates consistently on the Pareto frontier of the latency-accuracy skyline. 
Meanwhile, the head parallel load balancer itself reduces latency by 1.26$\times$.

\section{Background and Motivation}
\label{sec:background}

\begin{figure}[tp]
    \centering
    \includegraphics[width=0.9\linewidth]{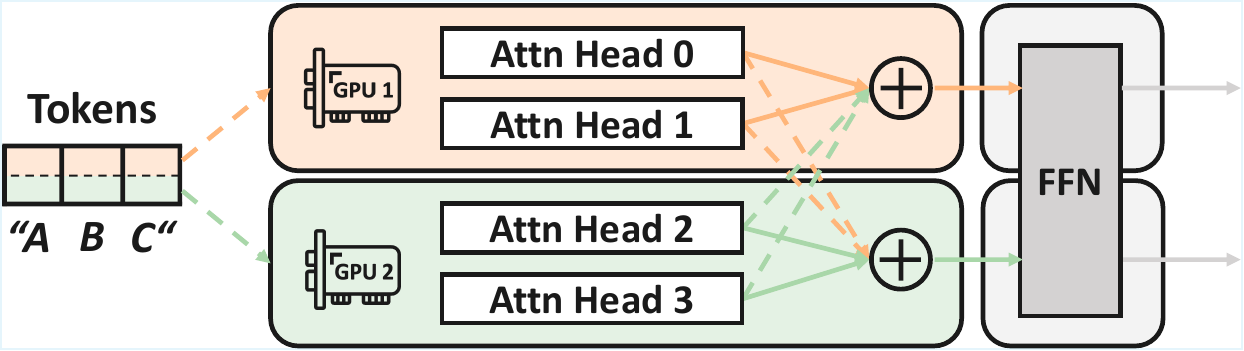}
    \caption{Illustration of the attention head parallelism.}
    \label{fig:bg_head_parallel}
\end{figure}

\subsection{Attention in LLM Inference}
\label{subsec:transformer-attention}

A critical operation of LLM inference is self-attention. 
Given an input of $n$ tokens, self-attention first transforms them into three matrices: query $Q$, key $K$, and value $V$. Using the query and key, the attention weight $A$ is derived and the output $O$ is computed as the weighted sum over the value $V$:

\begin{equation}
A = \text{Softmax}(\frac{Q\times K^T}{\sqrt{d}}), ~O =  A\times  V \label{eq:attn}
\end{equation}

In mainstream attention architectures such as Multi-Head Attention (MHA) and Group-Query Attention (GQA)~\cite{Ainslie2023gqa}, the attention operation is typically composed of multiple modules, each known as an attention head (e.g., 32 heads for the Llama-3.1-8B model).
Each head independently performs self-attention as in Eq.~\eqref{eq:attn} and captures diverse features from different subspaces. The results from all heads are then aggregated to yield the output.

Moreover, recent LLMs have demonstrated exceptional performance in processing \emph{long contexts}, supporting a wide range of applications like document understanding and repository-level code debugging~\cite{bairi2024codeplan, jimenez2023swe}. 
While crucial for capturing relationships within sequences, efficiently serving attention in long-context scenarios remains challenging. 
In the prefill phase, processing all input tokens at once incurs quadratic computational complexity, making the attention operation compute-bound. As an illustration, the attention operation accounts for over 70\% of the total latency when handling a 128K context length input in the Llama-3.1-8B model. In the decoding phase, key-value (KV) cache optimization reduces the complexity to linear, but the operation becomes memory-bound due to frequent memory accesses. In this work, we focus on optimizing the computation cost in the prefill phase, complementing existing studies that optimize decoding attention.

\subsection{Distributed Attention Deployment}
\label{subsec:dist-llm-serv}

Since attention has become a critical bottleneck, existing work has focused on optimizing attention serving through distributed model deployment strategies~\cite{zhu2025megascale, wang2025step,Zheng2024Efficient, Kwon2023efficient}. Given the differing computational characteristics between the attention and other operations, some studies have proposed attention-FFN disaggregation (AFD) deployment strategies~\cite{zhu2025megascale, wang2025step}. AFD makes optimizing attention serving efficiency as the sole objective possible. For the attention operation itself, current serving systems primarily optimize inference latency through tensor parallelism. As discussed in~\Cref{subsec:transformer-attention}, the multi-head nature of attention is inherently suitable for parallelization. Therefore, as shown in~\Cref{fig:bg_head_parallel}, leading serving systems~\cite{Zheng2024Efficient, Kwon2023efficient} typically employ head parallelism (HP) for attention to improve serving quality.

\begin{figure}[tp]
    \centering
    \includegraphics[width=0.85\linewidth]{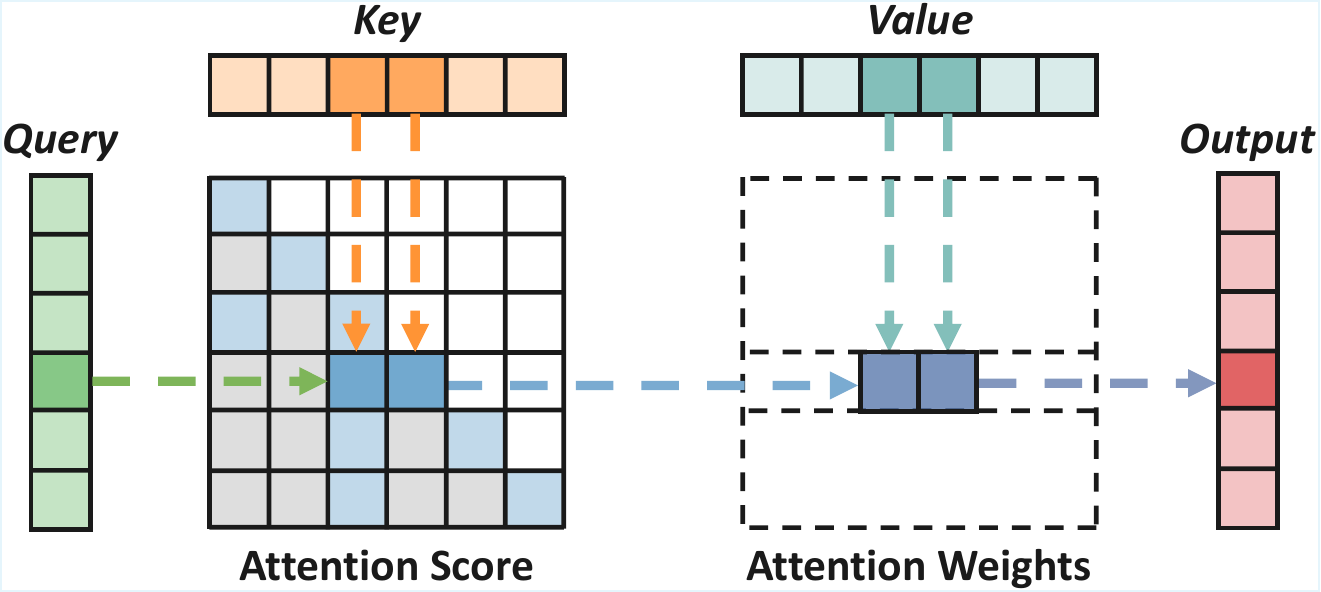}
    \caption{Illustration of the sparse attention workflow.}
    \label{fig:bg_sparse_attn}
\end{figure}

\subsection{Sparse Attention}
\label{subsec:sparse-attn}

Apart from the system-side distributed attention deployment, sparse attention, as an algorithm-side optimization technique, is also widely adopted for fast attention computation~\cite{xiao2023efficient,xu2025xattention,jiang2024minference,tang2024quest, zhang2024pqcache, liu2025retrievalattention, chen2024magicpig}.
Sparse attention leverages the inherent sparsity of the attention mechanism to alleviate the quadratic complexity. 
As illustrated in~\Cref{fig:bg_sparse_attn}, this sparsity arises naturally from the weighted average in attention computation, where only the value vectors associated with high attention weights contribute most significantly to the attention output. 
For example, Liu~et~al.~\cite{liu2025retrievalattention} reports that the top-1\% of the total tokens account for 89\% of the overall attention weight.
Ideally, if a proper number of most-significant query-key pairs are selected in attention calculation, the inference process can be substantially accelerated with negligible accuracy degradation.


For example, some works observe structure sparsity patterns of the attention map in prefill phase and accordingly optimize the quadratic computation overhead.
StreamingLLM~\cite{streamingllm} proposes to only compute the initial and recent tokens in the context window, and MInference~\cite{jiang2024minference} combines different sparsity patterns (e.g., local window, vertical-slash and block sparse) when selecting the token candidates for attention computation.
Despite with different selection principles, all those works rely on a critical hyperparameter---the token budget ($k$)---to specify the ultimate number of tokens that participate in attention computation on each attention head. 

\subsection{Cross-head Sparsity Heterogeneity} 
\label{subsec:head_hetero}

\begin{figure}[tp]
    \centering
    \begin{minipage}[b]{0.48\linewidth} 
        \centering
        \includegraphics[width=1\linewidth]{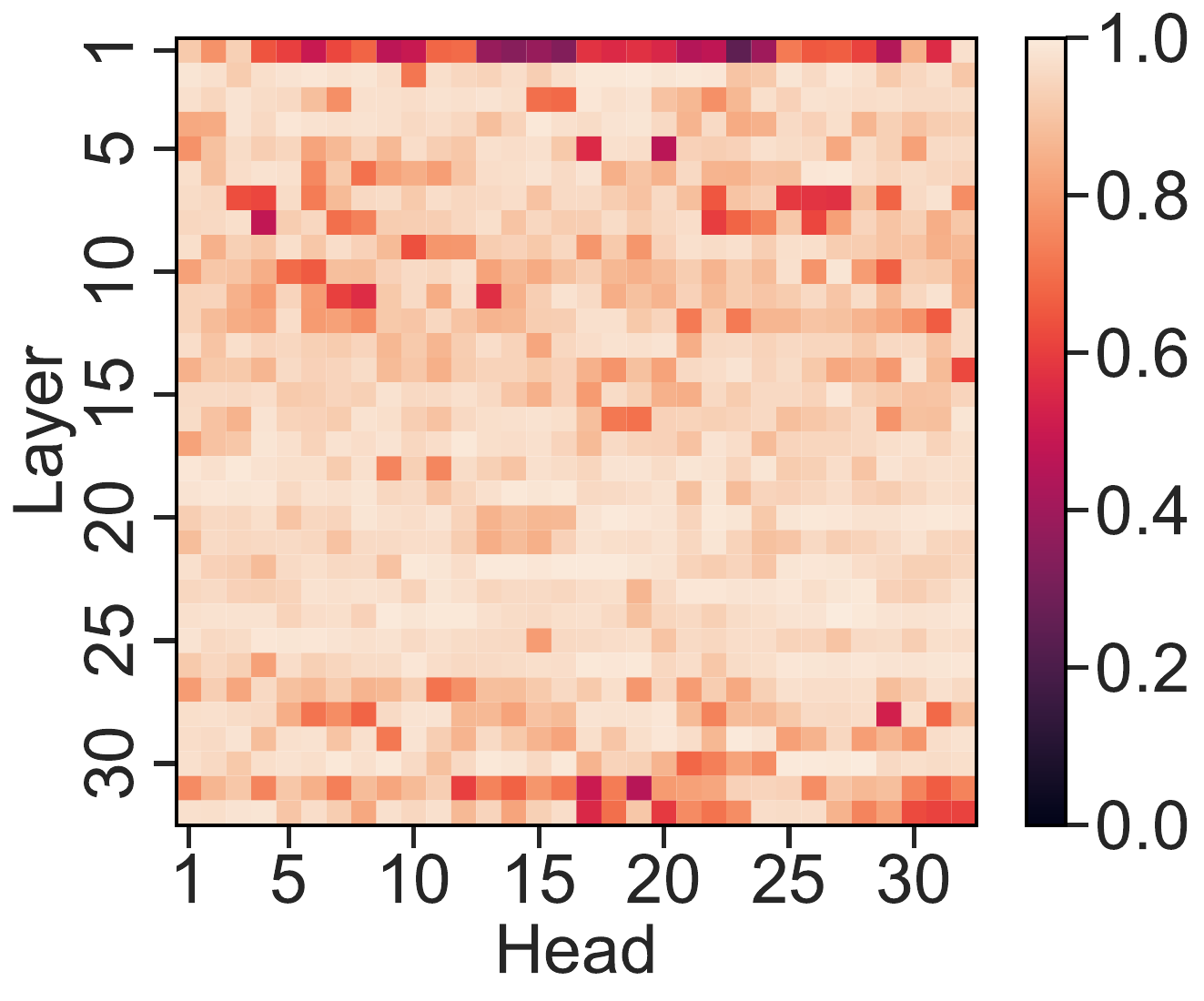}
        \caption{Heterogeneity of attention weight recovery ratio across different heads.}
        \label{fig:moti_fixed_budget}
    \end{minipage}
    \hfill 
    \begin{minipage}[b]{0.48\linewidth} 
        \centering
        \includegraphics[width=1\linewidth]{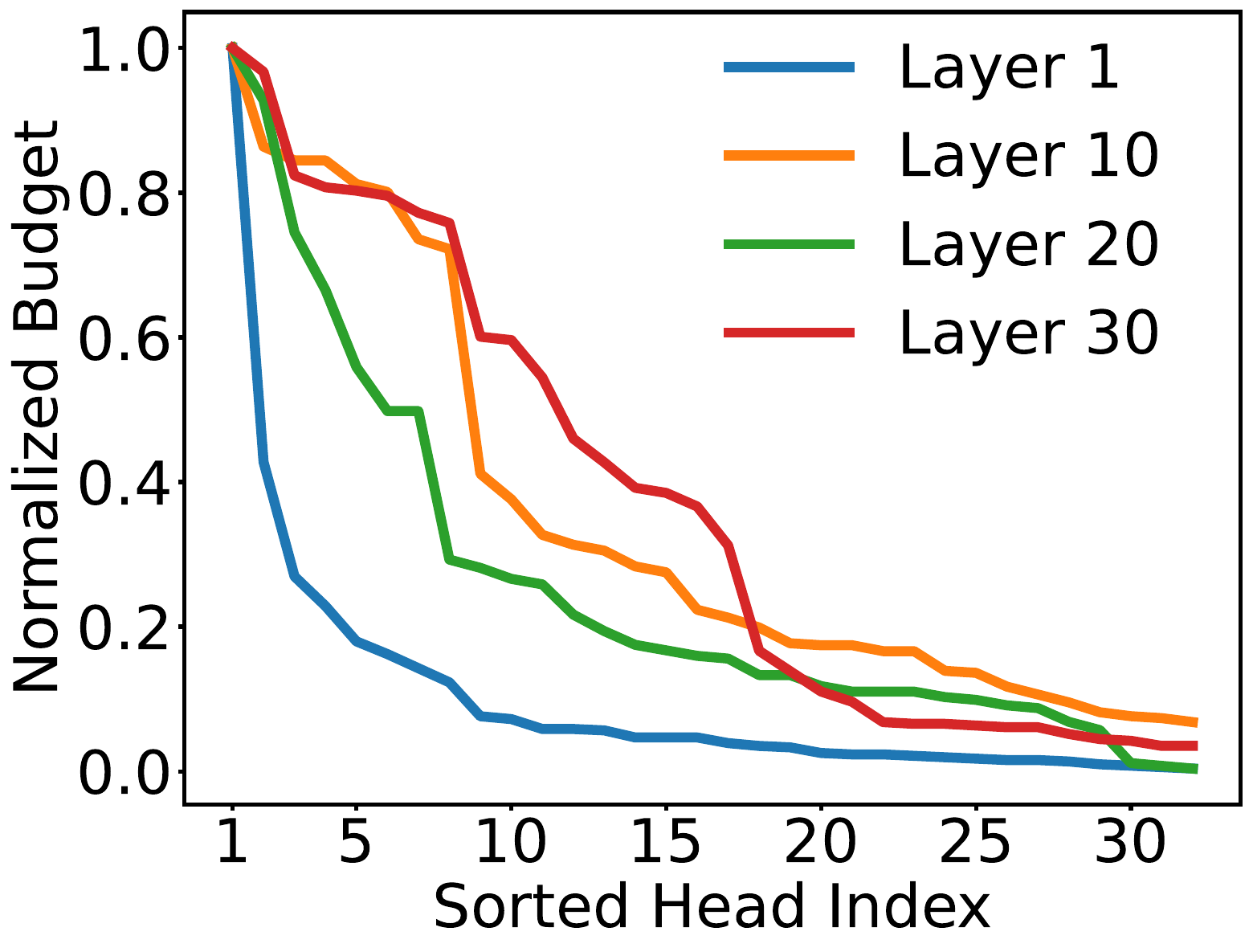}
        \caption{Normalized budget for each head across multiple layers. }
        \label{fig:moti_fixed_threshold}
    \end{minipage}
\end{figure}

While the above sparse-attention works enforce an identical token budget on each head, we find that different attention heads of an LLM model often exhibit heterogeneous sparsity characteristics.
That is, some attention heads require only a small number of tokens to recover high attention weight (i.e., $A$ in Eq.~\eqref{eq:attn}), while others do not.
To confirm, we quantify the attention sparsity by calculating the cumulative sum of attention weight of top-$k$ critical tokens. 
This cumulative sum, called recovery ratio, indicates how much of the full attention can be recovered using a small number of critical tokens (a higher recovery ratio reflects a larger sparsity). 
When processing a 32K context-length input (from PG-19~\cite{Rae2020Compressive}) using the Llama-3.1-8B model, we profile the recovery ratio for the top-$k$ (k=4096) tokens across different attention heads. 
As shown in~\Cref{fig:moti_fixed_budget}, there exists considerable heterogeneity in the recovery ratio of different attention heads.
Such heterogeneity is also corroborated by some recent works~\cite{Wu2025Retrieval, xiao2025duoattention}. 

Given the existence of cross-head sparsity heterogeneity, it would be suboptimal to enforce a uniform token budget $k$ on each attention head.
When $k$ is set too high, it leads to excessive budget allocation for sparser heads, resulting in redundant computations; conversely, when $k$ is too low, it allocates insufficient budget to less sparse heads, resulting in compromised accuracy.

Recently, some works have proposed top-$p$ strategy~\cite{xu2025xattention,lin2025twilight} to implicitly handle cross-head heterogeneity, where each attention head includes all the significant tokens whose cumulative attention weight exceeds a certain threshold $p$.
While this top-$p$ method can attain relatively good accuracy performance by adaptively adjusting the number of tokens in attention computation, it nonetheless has significant deficiencies. 
First, it incurs large analysis overheads to transfer the top-$p$ objective to the specific token candidates. Unlike top-$k$ methods, which only require ordinal information, the threshold-based top-$p$ method demands more precise (often inaccessible) online estimation of varying token importance.
Meanwhile, when allowing each head to have a distinct token budget, the attention computation amount across different heads---and further across different GPUs in HP---would be inconsistent. We sample the number of tokens required for each head across several layers at $p=0.9$, and the results are shown in~\Cref{fig:moti_fixed_threshold}. This demonstrates a significant variation in the budget required by different heads to meet the same threshold. Since the downstream FFN module cannot be launched before the completion of all the heads' attention computation, this would incur remarkable resource wastage at the module barriers. 

To summarize, while different attention heads have heterogeneous sparsity characteristics, it is however a non-trivial task to leverage such cross-head attention heterogeneity for efficient and accurate attention serving.
On the one hand, we need to precisely determine the token (key vector) candidates for sparsified attention computation on each head---so as to make substantial reduction on the overall computation amount.
On the other hand, we need to minimize the resource wastage caused by inconsistent computation loads on different GPU devices.
We address the two challenges in the next section.

\section{\sysname{} Design}
\label{sec:solution}

\subsection{Architecture Overview}
\label{subsec:architecture_overview}

\begin{figure}[tp]
    \centering
    \includegraphics[width=0.85\linewidth]{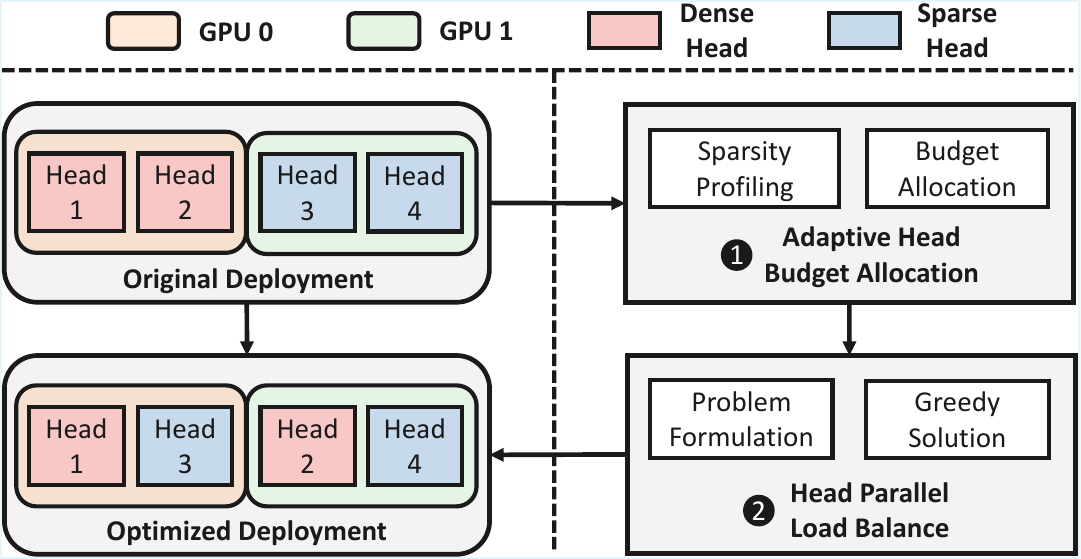}
    \caption{Architecture of \sysname{}.}
    \label{fig:solu_architecture}
\end{figure}

In this work, we propose \sysname{} (Sparsity-Aware Head-Parallel Load Balancing), a novel system-algorithm co-designed framework for efficient LLM attention serving. \sysname{} is tailored to strike an optimal balance between accuracy performance and computation efficiency. \Cref{fig:solu_architecture} illustrates the system architecture of \sysname{}, which comprises two core components:
\circleb{1}: 
Adaptive head budget allocation. It profiles the heterogeneous sparsity patterns of individual attention heads across diverse inputs via offline profiling, and then adaptively distributes computational budgets across the heads using a max–min optimization strategy.
\circleb{2}: 
Head parallel load balance. It formulates the head-to-device assignment as a multiway partitioning problem, and minimizes the inter-device load imbalance with a greedy algorithm.

\subsection{Adaptive Head Budget Allocation}
\label{subsec:adaptive_budget}

\phm{Stability of per-head sparsity characteristics.}
As detailed in \Cref{subsec:head_hetero}, cross-head sparsity heterogeneity demonstrates the necessity of allocating differentiated token budgets across heads to balance accuracy preservation and computation reduction. However, existing top-$p$ methods rely on online token significance analysis, which significantly exacerbates inference latency. 
Fortunately, our empirical findings reveal that, despite the variation in the sparsity of different heads, the relative sparsity of individual attention heads exhibits notable cross-request stability. 
In other words, the proportion of tokens required to recover a significant portion of the attention weight for each head is highly stable across different contexts or batches, even though the absolute number of tokens may vary.

\begin{figure}
\centering
\includegraphics[width=1\linewidth]{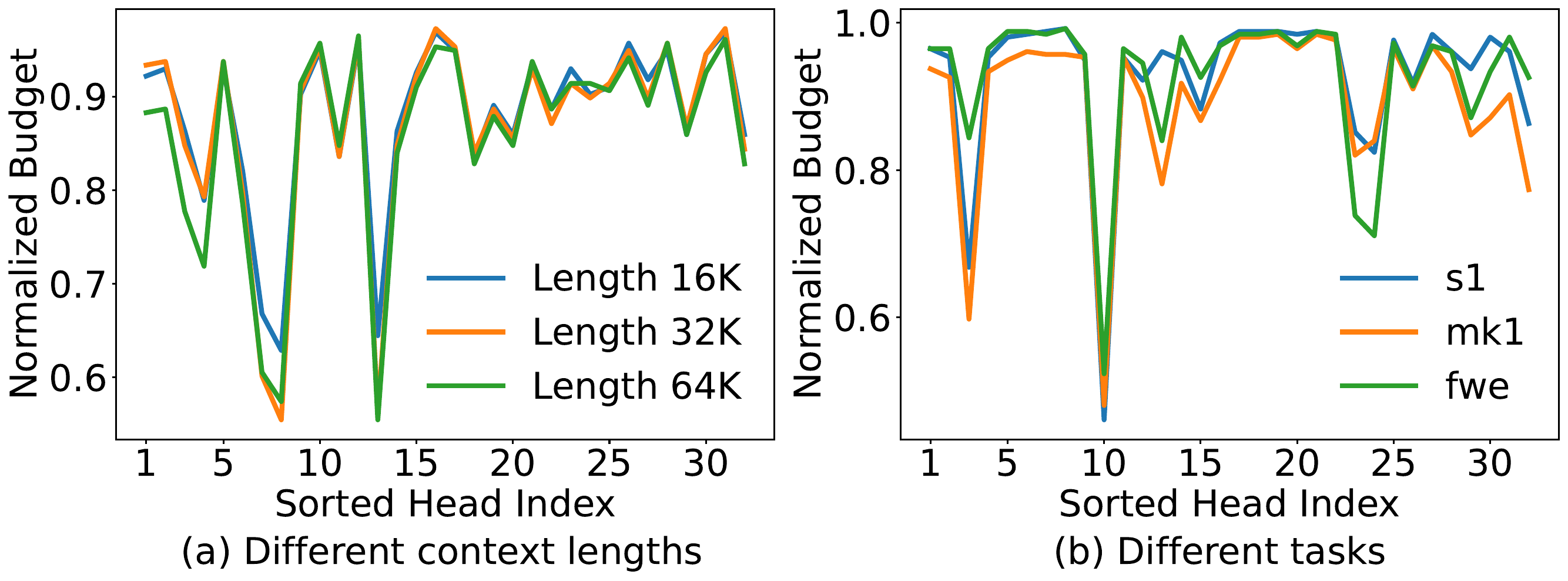}
\caption{Normalized budget required by different heads to achieve a fixed attention weight recovery ratio under inputs of varying context lengths and tasks.}
\label{fig:solu_stable}
\end{figure}

We perform experiments using the Llama-3.1-8B model on a series of calibration datasets~\cite{hsieh2024ruler}. Specifically, we carefully adjust the budget required for each attention head to reach a specified attention weight recovery ratio of 0.9 (across various context lengths and task domains, including classification, generation, and reasoning). The results presented in~\Cref{fig:solu_stable} demonstrate that the sparsity ratio for each attention head remains highly stable across different input datasets. This consistency across diverse tasks and context lengths suggests that the sparsity behavior of attention heads generalizes well, rather than being specific to a particular scenario. This finding is also corroborated by several existing studies~\cite{Wu2025Retrieval}.
Therefore, offline profiling on a calibration dataset can provide valuable insights for determining the sparsity level of each attention head and precomputing the optimal budget for each head.

\phm{Max-min sparsity budget shifting.}
We introduce an efficient algorithm for budget allocation across attention heads using the max-min strategy. 
Top-$p$ methods approximate the accuracy of full attention by including tokens with cumulative attention weights above a threshold. However, it is challenging to precisely find the best token candidates that in total yield the desired $p$ attention score,
making the top-$p$ method computationally inefficient and also inaccurate~\cite{lin2025twilight}. To achieve both efficiency and high accuracy, we propose an algorithm that approximates top-$p$ accuracy within a fixed computational amount, using a budget allocation strategy based on the max-min principle. The core idea is to iteratively adjust the budget allocation across attention heads, assigning more resources to heads with lower sparsity while reducing redundant computation for sparser heads. This approach strikes a balance between efficiency and accuracy, ensuring effective budget utilization while maintaining performance close to top-$p$.

\begin{figure}[tp]
    \centering
    \begin{minipage}[b]{0.48\linewidth} 
        \centering
        \includegraphics[width=\linewidth]{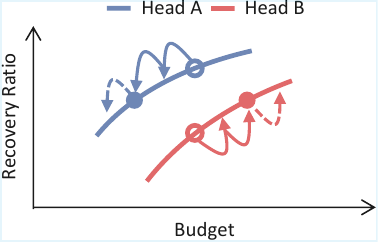}
        \caption{Illustration of the iterative max–min budget allocation process.}
        \label{fig:solu_max_min_algo}
    \end{minipage}
    \hfill 
    \begin{minipage}[b]{0.48\linewidth} 
        \centering
        \includegraphics[width=\linewidth]{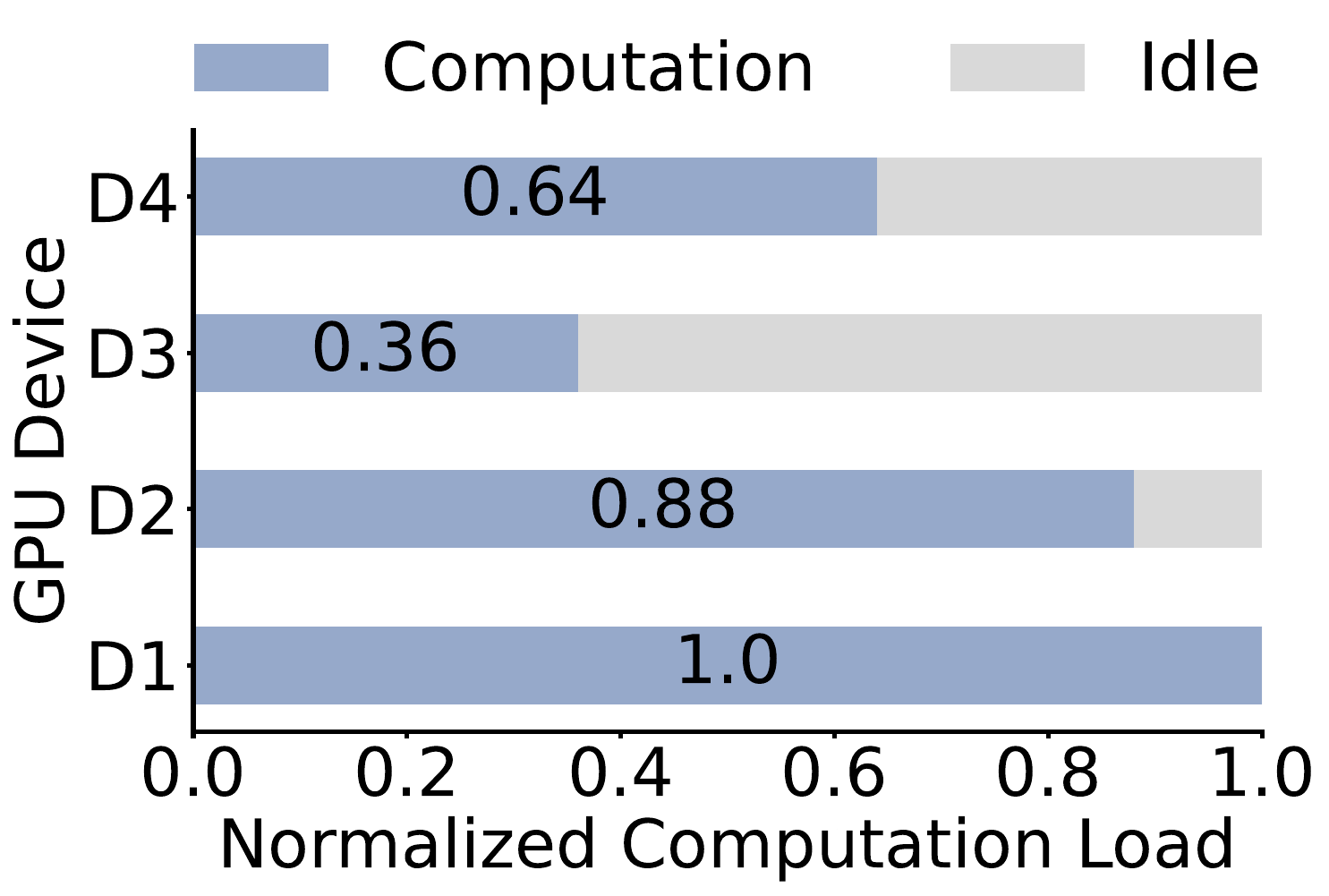}
        \caption{Load imbalance arising from naive head parallel attention deployment.}
        \label{fig:solu_max_min_load_imbalance}
    \end{minipage}
\end{figure}

The budget allocation algorithm proceeds in several stages. As illustrated in~\Cref{fig:solu_max_min_algo}, each attention head is first assigned an equal budget, while their attention-weight recovery ratios may differ (hollow circles). The algorithm then iteratively transfers budget from the head with the highest sparsity to the one with the lowest sparsity. This adjustment continues until one of the following conditions is reached: (i) additional reallocation no longer yield benefits (dashed line), indicating that the budget‐providing head has become the new minimum; or (ii) no further budget can be extracted because all attention heads have reached a predefined lower bound (typically a small value such as 128).

However, allocating different budgets to attention heads introduces cross-GPU load imbalance. Existing head-parallel (HP) methods assume that all attention heads have equal computational cost and thus assign them sequentially. Differentiated budget allocation breaks this assumption, leading to uneven computational loads.~\Cref{fig:solu_max_min_load_imbalance} shows the load imbalance observed when a specific layer of the Llama-3.1-8B model is naively distributed across 4 GPUs, with similar patterns appearing in other layers. The imbalance can be as high as $\textbf{2.78}\times$, highlighting the need for a more effective load balance strategy to ensure consistent computation times across devices and improve overall resource utilization.

\subsection{Head Parallel Load Balance}
\label{subsec:hplb}

In \Cref{subsec:adaptive_budget}, we design an adaptive budget allocation mechanism to strike an optimal trade-off between accuracy and efficiency. However, computational inconsistency across heads may lead to substantial GPU idleness, as faster GPUs are forced to wait for straggler devices to complete their computations.
Therefore, in this section, we first formulate the load balancing optimization problem and subsequently present an efficient greedy algorithm, Head Parallel Load Balance, to resolve cross-GPU load inconsistency.

\phm{Problem formulation.}
To address the load imbalance issue, we formulate an optimization problem to efficiently allocate attention heads across multiple devices. Given a set of attention heads $H = \{h_1, h_2, \dots, h_N\}$ with corresponding budgets $b_{h_1}, b_{h_2}, \dots, b_{h_N}$. Let $D$ represent the set of devices, and $H_d \in H$ be the subset of heads assigned to device $d\in D$. The optimization problem can be formulated as follows:

\begin{equation}
\begin{aligned}
\min_{\{H_d\}_{d \in D}} \quad & \mathcal{I} = \frac{\max_{d \in D} L_d}{\frac{1}{|D|} \sum_{d \in D} L_d}, \\
\text{s.t.} \quad & L_d = \sum_{h \in H_d} b_h, \quad \forall d \in D, \\
& \bigcup_{d \in D} H_d = H, \\
& H_{d_1} \cap H_{d_2} = \emptyset, \quad \forall d_1 \neq d_2, 
\end{aligned}
\end{equation}

where $L_d$ denotes the total load of the device $i$. Our objective is to minimize the load imbalance ratio $\mathcal{I}$, subject to the constraint that each head is assigned to exactly one device.

\begin{table*}[h]
    \centering
    \setlength{\tabcolsep}{2mm}
    \caption{Performance (\%) comparison of different models and methods on RULER. \sysname{} outperforms all other methods and maintains accuracy comparable to that of full attention across all models.
    }
    \label{tab:ruler_acc}
    \setlength{\tabcolsep}{6pt}
    \renewcommand{\arraystretch}{0.95}
    \small
    \begin{tabularx}{1\textwidth}{c|l|ccccccccccccc|c}
    \toprule
    & \textbf{Methods} & NS1 & NS2 & NS3 & MK1 & MK2 & MK3 & MV & MQ & VT & CWE & FWE & QA1 & QA2 & \textbf{Avg.} \\
    
    \midrule
    \multirow{5}{*}{\rotatebox{90}{\textit{Llama-3.1-8B}}} & Full Attention & 100.00 & 100.00 & 99.00 & 99.00 & 87.00 & 58.00 & 91.50 & 97.75 & 90.00 & 0.00 & 51.67 & 76.00 & 43.00 & 76.38 \\
    &StreamingLLM & 93.00 & 48.00 & 38.00 & 32.00 & 6.00 & 0.00 & 35.75 & 33.75 & 68.60 & 0.10 & 58.67 & 16.00 & 24.00 & 34.91 \\
    &MInference & 99.00 & 97.00 & 100.00 & 95.00 & 78.00 & 33.00 & 84.25 & 94.00 & 79.00 & 0.20 & 52.33 & 65.00 & 40.00 & 70.52 \\
    &XAttention & 100.00 & 100.00 & 99.00 & 97.00 & 84.00 & 44.00 & 92.50 & 97.50 & 89.20 & 0.20 & 49.33 & 56.00 & 44.00 & 73.29 \\
    &\cellcolor{green!5}\textbf{\sysname{}} & \cellcolor{green!5}100.00 & \cellcolor{green!5}99.00 & \cellcolor{green!5}100.00 & \cellcolor{green!5}95.00 & \cellcolor{green!5}86.00 & \cellcolor{green!5}63.00 & \cellcolor{green!5}86.75 & \cellcolor{green!5}96.50 & \cellcolor{green!5}86.80 & \cellcolor{green!5}0.50 & \cellcolor{green!5}58.67 & \cellcolor{green!5}72.00 & \cellcolor{green!5}42.00 & \cellcolor{green!5}\textbf{75.86} \\
    
    \midrule
    \multirow{5}{*}{\rotatebox{90}{\textit{Qwen2.5-7B}}} & Full Attention & 100.00 & 100.00 & 92.00 & 93.00 & 49.00 & 17.00 & 69.50 & 86.25 & 83.80 & 37.70 & 62.76 & 50.00 & 36.00 & 67.46 \\
    &StreamingLLM & 77.00 & 84.00 & 88.00 & 27.00 & 4.00 & 1.00 & 45.25 & 34.25 & 23.40 & 10.80 & 73.00 & 11.00 & 21.00 & 38.44 \\
    &MInference & 100.00 & 96.00 & 96.00 & 90.00 & 47.00 & 5.00 & 59.00 & 81.50 & 72.60 & 20.90 & 66.00 & 36.00 & 33.00 & 61.77 \\
    &XAttention & 100.00 & 100.00 & 95.00 & 84.00 & 22.00 & 13.00 & 69.00 & 90.75 & 75.60 & 37.50 & 64.00 & 35.00 & 39.00 & 63.45 \\
    &\cellcolor{green!5}\textbf{\sysname{}} & \cellcolor{green!5}100.00 & \cellcolor{green!5}100.00 & \cellcolor{green!5}96.00 & \cellcolor{green!5}94.00 & \cellcolor{green!5}59.00 & \cellcolor{green!5}11.00 & \cellcolor{green!5}67.00 & \cellcolor{green!5}90.50 & \cellcolor{green!5}70.60 & \cellcolor{green!5}30.10 & \cellcolor{green!5}63.00 & \cellcolor{green!5}38.00 & \cellcolor{green!5}40.00 & \cellcolor{green!5}\textbf{66.09} \\
    
    \midrule
    \multirow{5}{*}{\rotatebox{90}{\textit{Qwen2.5-72B}}} & Full Attention & 100.00 & 100.00 & 98.00 & 100.00 & 82.00 & 28.00 & 96.50 & 100.00 & 100.00 & 92.80 & 86.67 & 54.00 & 50.00 & 83.69 \\
    &StreamingLLM & 10.00 & 26.00 & 92.00 & 26.00 & 6.00 & 0.00 & 9.50 & 48.50 & 4.40 & 3.00 & 87.50 & 18.00 & 24.00 & 27.30 \\
    &MInference & 100.00 & 100.00 & 100.00 & 92.00 & 74.00 & 38.00 & 90.50 & 99.50 & 85.20 & 74.00 & 86.81 & 42.00 & 54.00 & 79.69 \\
    &XAttention & 100.00 & 100.00 & 98.00 & 96.00 & 58.00 & 20.00 & 93.00 & 98.00 & 96.00 & 88.80 & 87.50 & 46.00 & 58.00 & 79.95 \\
    
    &\cellcolor{green!5}\textbf{\sysname{}} & \cellcolor{green!5}100.00 & \cellcolor{green!5}100.00 & \cellcolor{green!5}100.00 & \cellcolor{green!5}98.00 & \cellcolor{green!5}80.00 & \cellcolor{green!5}48.00 & \cellcolor{green!5}92.00 & \cellcolor{green!5}100.00 & \cellcolor{green!5}82.80 & \cellcolor{green!5}79.00 & \cellcolor{green!5}85.42 & \cellcolor{green!5}34.00 & \cellcolor{green!5}48.00 & \cellcolor{green!5}\textbf{80.56} \\
    \bottomrule
    
    \end{tabularx}
\end{table*}

\phm{A greedy heuristic for efficient problem solving.}
This attention head deployment problem is essentially an NP-hard multiway partitioning problem~\cite{cong1998multiway}.
Although exact algorithms such as dynamic programming guarantee an optimal solution, their state space grows exponentially with the number of devices $|D|$ and the upper budget bound $L$, with a complexity of $O(N \cdot L^{|D|-1})$. This makes them computationally infeasible when either $D$ or $L$ is large. Therefore, we employ an efficient greedy algorithm to solve this problem. The algorithm begins by sorting attention heads in descending order of their budgets. Each attention head is then assigned to the device with the currently smallest load. This process continues until all heads have been allocated. The greedy approach minimizes the load imbalance by prioritizing the allocation of the most demanding heads first, ensuring a balanced load distribution. 
This algorithm has a relatively low time complexity of $O(N \log N + N \log K)$, making it suitable for adoption in practice.

\section{Implementation}
\label{sec:implementation}

\sysname{} is implemented with 1.6K lines of Python code on top of an existing sparse attention framework, MInference~\cite{jiang2024minference}. Since \sysname{} focuses on adaptive budget allocation for different attention heads, we build upon the sparse pattern kernels from MInference. \sysname{} extends the adaptive head budget allocation algorithm and incorporates a sparsity-aware attention head deployment strategy to achieve load balance in head-parallel scenario.
\section{Evaluation}
\label{sec:eval}

In this section,  we comprehensively evaluate \sysname{}, comparing it with full attention and other state-of-the-art sparse attention methods on different models and tasks.

\begin{figure*}[h!]
    \centering
    \includegraphics[width=1\linewidth]{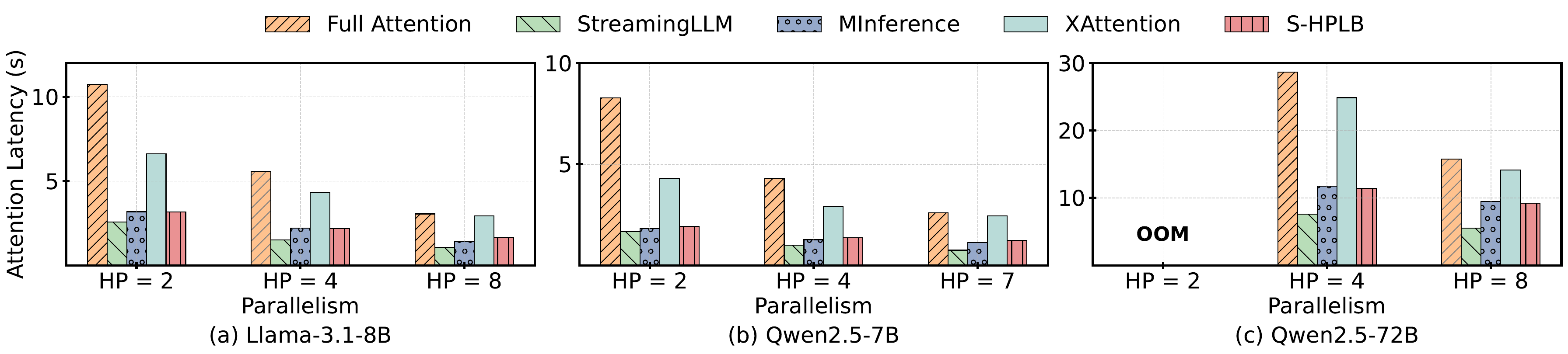}
    \caption{Latency of attention serving across models with 128K input. \sysname{} outperforms the top-$p$ method and achieves comparable performance to the top-$k$ method.}
    \label{fig:e2e_latency}
\end{figure*}

\begin{figure}
\centering
\includegraphics[width=1\linewidth]{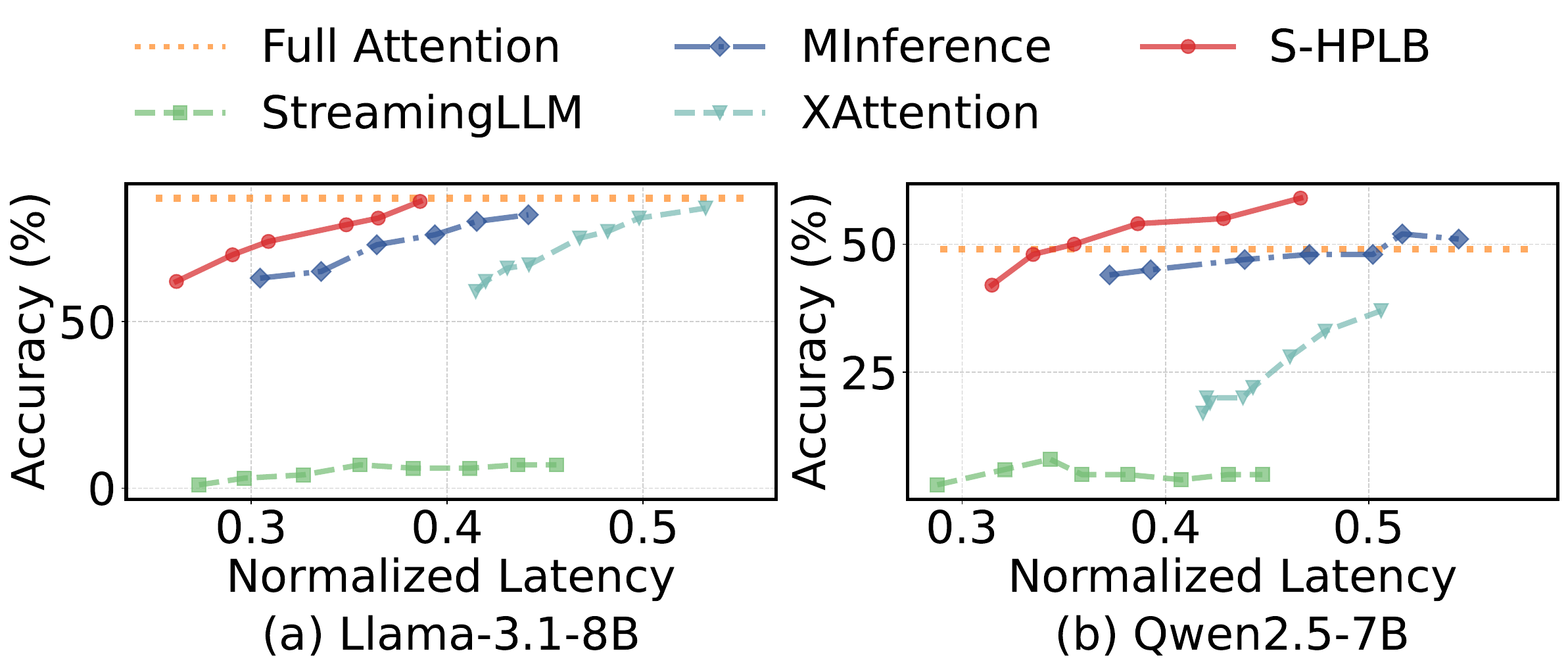}
\captionof{figure}{Accuracy-latency trade-off (Skyline) across various methods on two models.}
\label{fig:accuracy_latency_skyline}
\end{figure}

\subsection{Experimental Setup}

\phm{Models and benchmark.}
We evaluate \sysname{} primarily on three leading open-source LLMs: Llama-3.1-8B~\cite{llama-3.1-8B}, Qwen2.5-7B~\cite{qwen2.5}, and Qwen2.5-72B~\cite{qwen2.5-72B}. All models support a context window of up to 128K tokens. Moreover, we include the Llama-3-8B-262K model~\cite{llama-3-8B-262k} in our evaluation to assess efficiency and scalability in scenarios with longer contexts.
We evaluate model accuracy on the widely-used long-context benchmark, RULER~\cite{hsieh2024ruler}, which is designed to challenge sparse attention methods. RULER consists of four categories and thirteen tasks, including retrieval, multi-hop tracing, aggregation, and QA, with context lengths of up to 128K.

\phm{Hardware testbed.} 
All experiments were conducted on a single server equipped with an Intel Xeon Platinum 8369B CPU and eight NVIDIA A100 80GB GPUs connected via NVLink. The server runs CentOS 7 with CUDA 12.4 and torch 2.6.0.

\phm{Baselines.}
We compare \sysname{} against several strong baselines representing the state-of-the-art in efficient LLM inference. For full attention baseline we run FlashAttention~\cite{dao2022flashattention}. For sparse attention, we select representative methods from two categories:
(1) Top-$k$ methods: StreamingLLM~\cite{xiao2023efficient} and MInference~\cite{jiang2024minference}. These methods rely on a predefined computational budget, but differ in how the budget is allocated: StreamingLLM adopts an initial and tail local window, while MInference employs  three distinct sparse attention patterns.
(2) Top-$p$ method: XAttention~\cite{xu2025xattention}. This method dynamically determines the budget at runtime to ensure the cumulative attention weight meets a predefined threshold. We set the hyperparameter for all methods according to original papers.

\phm{Metrics.} We evaluate \sysname{} based on two primary metrics: model accuracy and average attention serving latency (in time to first token). These metrics reflect \sysname{}'s performance from both the algorithmic and system perspectives.

\subsection{End-to-End Results}
This section provides a comparative analysis of various attention mechanisms, aiming to demonstrate the accuracy and efficiency of \sysname{} in comparison to existing methods.

\phm{Accuracy.}~\Cref{tab:ruler_acc} shows the model accuracy of various methods on the RULER benchmark. \sysname{} consistently outperforms the other sparse-attention methods and matches\footnote{
Not that for certain tasks (e.g., MK3 on Llama3.1-8B, and also for some points in later \Cref{fig:accuracy_latency_skyline}), \sysname{} achieves even higher accuracy than the full attention mechanism. This improvement can be attributed to \sysname{}'s ability to filter out noise and help the model concentrate on critical tokens. Other sparse attention studies have also reported this phenomenon \cite{chen2025retroinfer}.
} the average accuracy of full attention. \sysname{} experiences only a drop of 0.52\%/1.37\%/3.13\% compared to full attention on Llama3.1-8B/Qwen2.5-7B/Qwen2.5-72B, respectively. 
When compared to the best-performing sparse attention baseline XAttention, \sysname{} achieves accuracy improvements of 2.57\%/2.94\%/0.61\% on each model, respectively. 
Intuitively, the top-$p$ method should achieve the best accuracy. However, in practice, due to the inaccurate and computationally expensive estimation of attention weights, its performance is suboptimal.

\phm{Attention latency.}
We evaluate the attention serving latency across different models and degrees of parallelism under a 128K context length.
As shown in~\Cref{fig:e2e_latency}, \sysname{} achieves $3.39\times$/$4.27\times$/$3.31\times$ lower latency than full attention on the three evaluated models.
In comparison with sparse attention baselines, \sysname{}  reduces latency by $2.09\times$/$2.22\times$/$2.88\times$ relative to XAttention. 
Benefited from head-parallel load balancing, \sysname{} also attains latency on par with top-$k$ methods.

\subsection{Sensitivity Study}
For a better understanding of \sysname{}, we further analyze how varying the configurations impacts overall performance. For top-$k$-based methods, we vary the uniform budget $k$, while for top-$p$-based methods, we adjust the threshold $p$. For \sysname{}, we modify the total budget across all heads. These settings directly control the sparsity level.
We evaluate the performance of all methods on the challenging MK2 task from the RULER benchmark~\cite{hsieh2024ruler} using two different models.~\Cref{fig:accuracy_latency_skyline} shows the curves between latency (normalized by the latency of full attention) and accuracy for various methods. Compared to other baselines, \sysname{} consistently occupies a more favorable position on the Pareto frontier, offering a better trade-off between accuracy and efficiency.

\subsection{Ablation Study}
\begin{figure}
\centering
\includegraphics[width=1\linewidth]{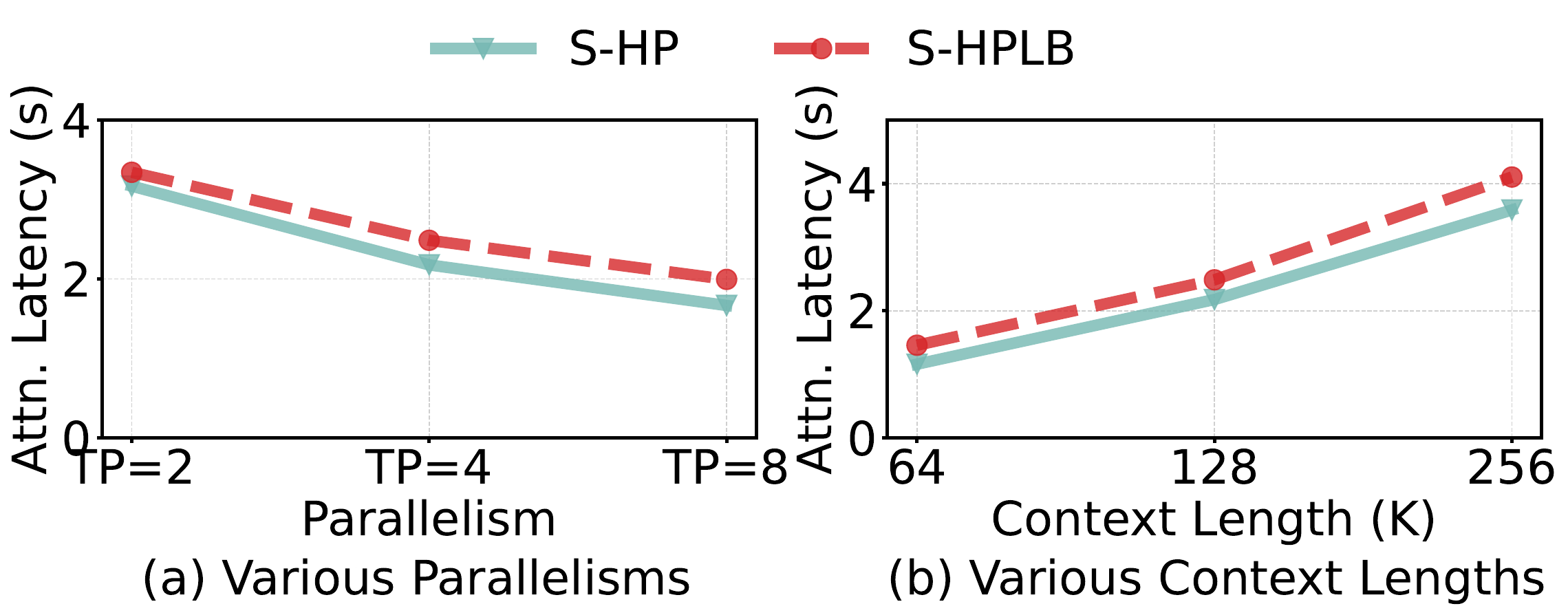}
\captionof{figure}{Impact of head parallel load balance under different parallelism degrees and context lengths.}
\label{fig:ablation_lb}
\end{figure}

To check the effectiveness of our head parallel load balancer, we experiment respectively with and without the load balancer.
We measure the attention latency across different degrees of parallelism with a fixed context length of 128K using the Llama-3.1-8B model. Additionally, we evaluate latency across different context lengths with a parallelism degree of 4.
As shown in \Cref{fig:ablation_lb}, the load balancer consistently reduces latency across all configurations. 
Across different parallelism degrees and context lengths, it achieves a latency reduction of up to $1.19\times$ and $1.26\times$, respectively.

\section{Related Work}
\label{sec:discussion}

\phm{Attention acceleration.} 
Recent research has explored various strategies to improve the efficiency of attention mechanisms~\cite{dao2022flashattention,Kwon2023efficient,gu2024mamba,Angelos2020Transformers}. System optimization techniques include FlashAttention~\cite{dao2022flashattention}, which accelerates attention computation by optimizing memory access patterns, and PagedAttention~\cite{Kwon2023efficient}, which addresses memory fragmentation through a paged management approach.
On the other hand, algorithmic optimization has been explored in several works~\cite{gu2024mamba, Angelos2020Transformers}, which propose alternative efficient attention methods.
These efforts are complementary to \sysname{}, which further enhances attention serving efficiency.

\section{Conclusion}
\label{sec:conlucsion}

This paper presents \sysname{}, a system-algorithm co-designed mechanism to improve the attention serving efficiency while maintaining high accuracy. 
\sysname{} leverages the sparsity stability of attention heads and introduces a novel adaptive head budget allocation algorithm. It further mitigates tail latency stemming from cross-GPU load inconsistency by employing head parallel load balance. Experimental results show that \sysname{} achieves an optimal trade-off between latency and accuracy, consistently operating on the Pareto frontier compared to the state-of-the-art sparse attention baselines.


\clearpage
\bibliographystyle{ACM-Reference-Format}
\bibliography{10-ref}


\end{document}